\documentclass[%
 reprint,
superscriptaddress,
 amsmath,amssymb,
 aps,prl
]{revtex4-1}
\usepackage{ulem}
\usepackage{graphicx}
\usepackage{dcolumn}
\usepackage{bm}
\usepackage{color}

\begin{document}

\title{Vortex multistability and Bessel vortices in polariton condensates}


\author{Xuekai Ma}
\affiliation{Department of Physics and Center for Optoelectronics and Photonics Paderborn (CeOPP), Universit\"{a}t Paderborn, Warburger Strasse 100, 33098 Paderborn, Germany}

\author{Stefan Schumacher}
\affiliation{Department of Physics and Center for Optoelectronics and Photonics Paderborn (CeOPP), Universit\"{a}t Paderborn, Warburger Strasse 100, 33098 Paderborn, Germany}
\affiliation{College of Optical Sciences, University of Arizona, Tucson, AZ 85721, USA}

\begin{abstract}

Vortices are topological objects formed in coherent nonlinear systems. As such they are studied in a wide number of physical systems and promise applications in information storage, processing, and communication. In semiconductor microcavities, vortices in polariton condensates can be conveniently created, studied, and manipulated using solely optical means. For non-resonant excitation with a ring-shaped pump a stable vortex can be formed, leading to  bistability with left- and right-handed vorticity. In the present work we report on a much richer vortex multistability, with optically addressable vortices with topological charges $m=\pm1$, $\pm2$, and $\pm3$, all stable for the same system and excitation parameters. This unusual multistable behavior is rooted in the inherent nonlinear feedback between reservoir excitations and condensate in the microcavity. For larger radius of the ring shaped pump we also find a Bessel vortex with its characteristic spiralling phase in the high density region and pronounced self-stabilization ability. Our theoretical results open up exciting possibilities for optical manipulation of vortex multiplets in a compact semiconductor system.
\end{abstract}


\maketitle

\textit{Introduction} -- Bistability and multistability are nonlinear phenomena observed in many physical systems such as magnetic systems~\cite{Sessoli-Nature-1993,Fujita-Science-1999}, semiconductors~\cite{Gibbs-APL-1979,Baas-PRA-2004}, atomic condensates~\cite{Treutlein-PRL-2007}, and nonlinear optical systems~\cite{Soljacic-PRE-2002,Boyd-Book-2003}. In a bistable or multistable nonlinear system, the solution space for a given set of parameters contains more than one stable state. For its potential use in all-optical switches, all-optical logic elements, optical transistors, and optical memory elements, optical bistability has been widely studied in a number of different system, including optical fibers~\cite{Fraile-OL-1991,Lyons-APL-1995}, photonic crystals~\cite{Yanik-APL-2003,Notomi-OE-2005}, and microcavities~\cite{Gibbs-APL-1979,Baas-PRA-2004}.

In the past decade, nonlinear optical physics with exciton polaritons in quantum well (QW) semiconductor microcavities have attracted a lot of attention. The fundamental optical excitations in this system are composed of QW excitons and cavity photons [cf. Fig.~1(a)]. Thanks to their photonic part, polaritons can be optically excited and probed, their matter part leads to pronounced optical nonlinearities. Making use of their condensed matter environment, coherent polariton ensembles can be efficiently created by off-resonant pumping into higher energy states of the semiconductor material [cf. Fig.~1(a)]. For elevated excitation densities and sufficient sample quality, subsequent relaxation and polariton-polariton scattering lead to accumulation of polaritons at the bottom of the lower-polariton branch. The resulting condensed polariton system then shows macroscopic cohererence \cite{Deng-science-2002,Kasprzak-nature-2006} even up to room temperature~\cite{Christopoulos-2007-prl,Christmann-2008-apl,Baumberg-2008-prl,Plumhof-2013-nm}. Besides lasing and condensation, these also include modulational instability~\cite{Borgh-PRB-2010,Luk-PRB-2013,Werner-PRB-2014,Liew-PRB-2015}, soliton and vortex formation~\cite{Egorov-2009-prl,Sich-2012-nph,Yulin-2008-pra,Grosso-2012-prb,Cilibrizzi-2014-prl,Ma-2017-prl}, and optical bistability~\cite{Baas-PRA-2004,Bajoni-PRL-2008,Liew-PRB-2010,Ballarini-NC-2013,Kyriienko-PRL-2014,Tan-PRB-2018}.

Vortices with their characteristic topological phase distribution can be created in a controlled manner in polariton systems using broad optical pumps~\cite{Lagoudakis-NatPhys-2008,Lagoudakis-Science-2009}, optically-induced two-dimensional parabolic potentials~\cite{Sigurdsson-prb-2014,Ma-prb-2016}, chiral polaritonic lenses~\cite{Dall-PRL-2014}, or ring-shaped intensity profiles~\cite{Ma-prb-2017}. A vortex with a given winding number has two possible topological charges, corresponding to clockwise or counter-clockwise rotation, which can be regarded as a type of bistability. However, this kind of bistability is trivial as the two vortex states have the same profile, the same existence and stability regions, and the same winding number; only the topological charges are opposite.

\begin{figure} 
\includegraphics[width=1\columnwidth]{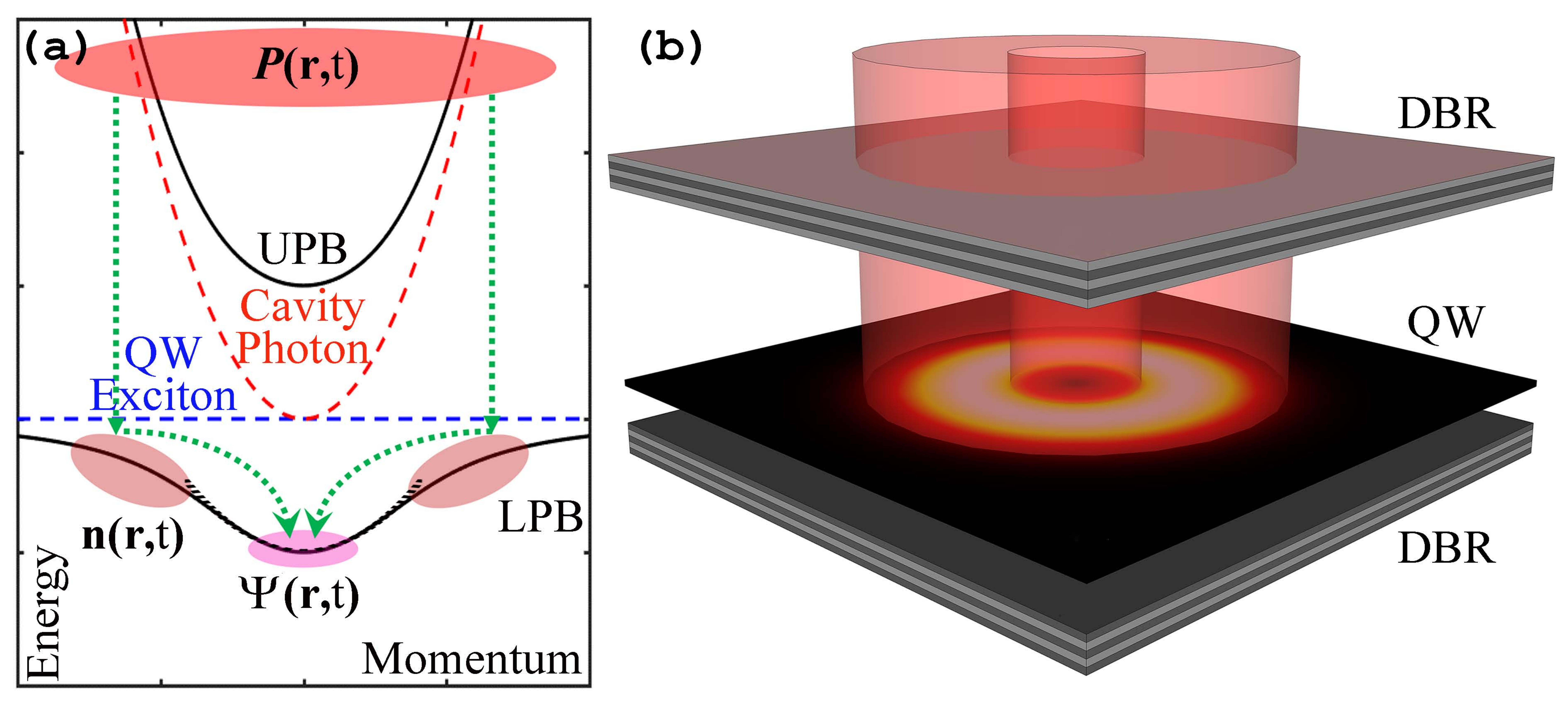}
\caption{(Color online) (a) Dispersions of bare QW exciton, cavity photon, and lower (LPB) and upper polariton branches (UPB). Off-resonant pumping and subsequent stimulated scattering from incoherent reservoir $n(\mathbf{r},t)$ to coherent condensate $\Psi(\mathbf{r},t)$ on the LPB in parabolic approximation are indicated. (b) Sketch of the planar semiconductor microcavity. A QW is placed between two distributed Bragg reflectors (DBRs) at the antinode of the cavity photon mode. An off-resonant ring pump is used to create and trap a polariton condensate in the center of the ring.}\label{microcavity}
\end{figure}


In this Letter, we report on a nontrivial vortex multistability in a polariton condensate. For nonresonant excitation with a continuous-wave (CW) ring-shaped pump [cf. Fig.1(b)], pump-induced excitations act as an incoherent source for the condensate, and at the same time provide an external potential~\cite{Schmutzler-prb-2015} trapping the coherent condensate in the center of the ring. 
Due to the stimulated scattering of excitations from the reservoir into the condensate, there is a pronounced feedback between condensate and reservoir. Even once a stationary solution is reached, the spatial shape of the reservoir is modified. We demonstrate that this feedback mechanism reshaping the reservoir, allows stabilization of vortices with different winding numbers for the same pump. We also demonstrate that switching between different vortex states can be achieved with an additional coherent light pulse carrying the same orbital angular momentum (OAM) as the target vortex state. For ring-shaped pumps with larger diameter we find the formation of vortices with a spiralling phase in the high-density region of the vortex. After switching off the pump source, as the density decays a Bessel vortex (nonradiation $J_1$ Bessel mode) is formed showing significant self-localization and reduced dispersion even at reduced densities \cite{Chong-NP-2010, Alexandrescu-PRA-2006}.

\textit{Model} -- The dynamics of the polariton condensate formed at the bottom of the lower-polariton branch [cf. Fig.~\ref{microcavity}(a)] is described by a mean-field driven-dissipative Gross-Pitaevskii (GP) model, coupled to the density of an incoherent reservoir~\cite{Wouters-prl-2007}:
\begin{equation}\label{e1}
\begin{aligned}
i\hbar\frac{\partial\Psi(\mathbf{r},t)}{\partial t}&=\left[-\frac{\hbar^2}{2m_{eff}}\nabla_\bot^2-i\hbar\frac{\gamma_c}{2}+g_c|\Psi(\mathbf{r},t)|^2 \right.\\
&+\left.\left(g_r+i\hbar\frac{R}{2}\right)n(\mathbf{r},t)\right]\Psi(\mathbf{r},t)+E(\mathbf{r},t)\,,
\end{aligned}
\end{equation}
\begin{equation}\label{e2}
\frac{\partial n(\mathbf{r},t)}{\partial t}=\left[-\gamma_r-R|\Psi(\mathbf{r},t)|^2\right]n(\mathbf{r},t)+P(\mathbf{r},t)\,.
\end{equation}
Here $\Psi(\mathbf{r},t)$ is the coherent polariton field and $n(\mathbf{r},t)$ is the reservoir density. $m_{eff}=10^{-4}m_e$ is the effective mass of polaritons at the bottom of the lower branch ($m_e$ is the free electron mass). Due to the finite lifetime of cavity polaritons, the condensate has a decay rate $\gamma_c=0.08\,\mathrm{ps^{-1}}$, and the reservoir decays with $\gamma_r=1.5\gamma_c$ \cite{Roumpos-NatPhys-2011}. The polariton condensate is replenished by the coupling to the reservoir density $n(\mathbf{r},t)$ with $R=0.01\,\mathrm{ps^{-1}\mu m^2}$, while the reservoir is excited by the incoherent pump $P(\mathbf{r},t)$. The condensate can directly be excited with a coherent light field $E(\mathbf{r},t)$. The interaction strength between polaritons is given by $g_c=3\times10^{-3}\,\mathrm{meV\mu m^2}$ and between polaritons and reservoir by $g_r=2g_c$. We note that we chose parameters for typical GaAs based microcavity systems. Even in this material system polariton lifetimes can range from a few picoseconds~\cite{
Deng-PNAS-2003,Love-PRL-2008,Wertz-NP-2010} to several hundreds of picoseconds~\cite{
Nelsen-PRX-2013,Steger-Optica-2015,Sun-PRL-2017}. In other materials also interaction strengths and consequently nonlinearities can be significantly different, with typical interaction strenghts of 1-10 $\mu$eV $\mu$m$^2$~\cite{Roumpos-NatPhys-2011,Ferrier-PRL-2011,Sanvitto-NM-2016} in inorganic materials and on the order of $10^{-3}$ $\mu$eV $\mu$m$^2$~\cite{Sanvitto-NM-2016,Daskalakis-PRL-2015,Lerario-NP-2017} in some organic materials.

\begin{figure} 
\includegraphics[width=1\columnwidth]{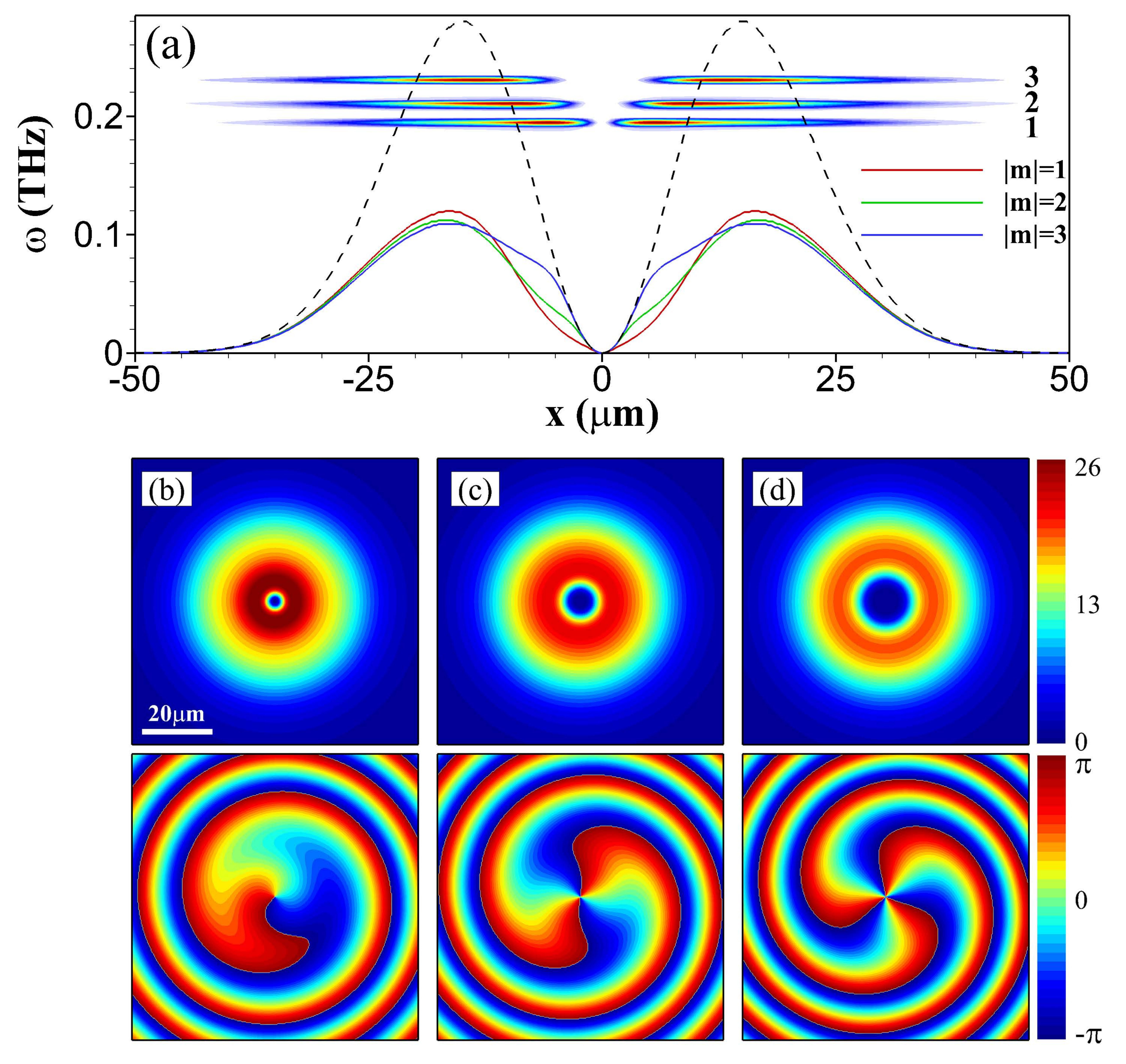}
\caption{(Color online) Multistability of vortices. (a) Spectra of multistable vortices with topological charges $|m|=1$, $2$, and $3$. Solid curves represent a cross section of the reservoir-induced potentials, given by $g_r n$, for the different topological charges. The dashed line represents the reservoir-induced potential with the condensation process switched off, i.e., $R=0$ in Eq. \eqref{e2}. (b)-(d) Distributions of densities and phases of vortices with topological charges (b) $m=1$, (c) $m=2$, and (d) $m=3$ for a pump with $P_0=10$ $\mathrm{ps^{-1}\mu m^{-2}}$ and $w=15$ $\mu $m.}\label{profiles}
\end{figure}

To create a vortex, a ring-shaped CW pump with $P(\mathbf{r})=P_0 \frac{\mathbf{r}^2}{w^2} e^{-\mathbf{r}^2/w^2}$ is used
with the radius of the ring, $w$ . The spatial profile of the pump is transferred to the excitation reservoir $n(\mathbf{r},t)$. 

\textit{Multistability} -- In previous study it was shown that a pump with a fixed ring shape supports one vortex state with a certain winding number~\cite{Ma-prb-2016}. However, here we find that the same pump can support vortices with different winding numbers [Fig.~\ref{profiles}] for a case with three different solutions. Which stationary state the system assumes depends on initial conditions. The spectra included in Fig.~\ref{profiles}(a) show that all condensates are squeezed to the higher energy states, instead of being trapped in the potential~\cite{Sigurdsson-prb-2014}, because of the strong feedback from the condensate. We note that in the numerical simulations this type of vortex multistability is generally more easily observed when the loss $\gamma_r$ of the reservoir is smaller than the condensation term $R|\Psi|^2$ depleting the reservoir. In this case for different condensate solutions, the reservoir experiences significantly different reshaping. Figure~\ref{profiles}(a) (dashed line) shows the unmodified external potential seen by the condensate for condensation rate $R=0$:
\begin{equation}\label{origPotential}
\omega(\mathbf{r})=\frac{g_r}{\hbar}n(r)=\frac{g_r}{\hbar}\frac{P(\mathbf{r})}{\gamma_r}.
\end{equation}
The other lines show the external potential seen by the condensate once the vortices have formed. In these cases the external potential trap is significantly modified by the presence of the vortex. For different winding numbers, the vortices have different radii and frequencies, which results in different reshaping of the potential. The effective radius of the vortex core and frequency of the vortex increase with increasing winding number. 

\begin{figure} 
\includegraphics[width=1\columnwidth]{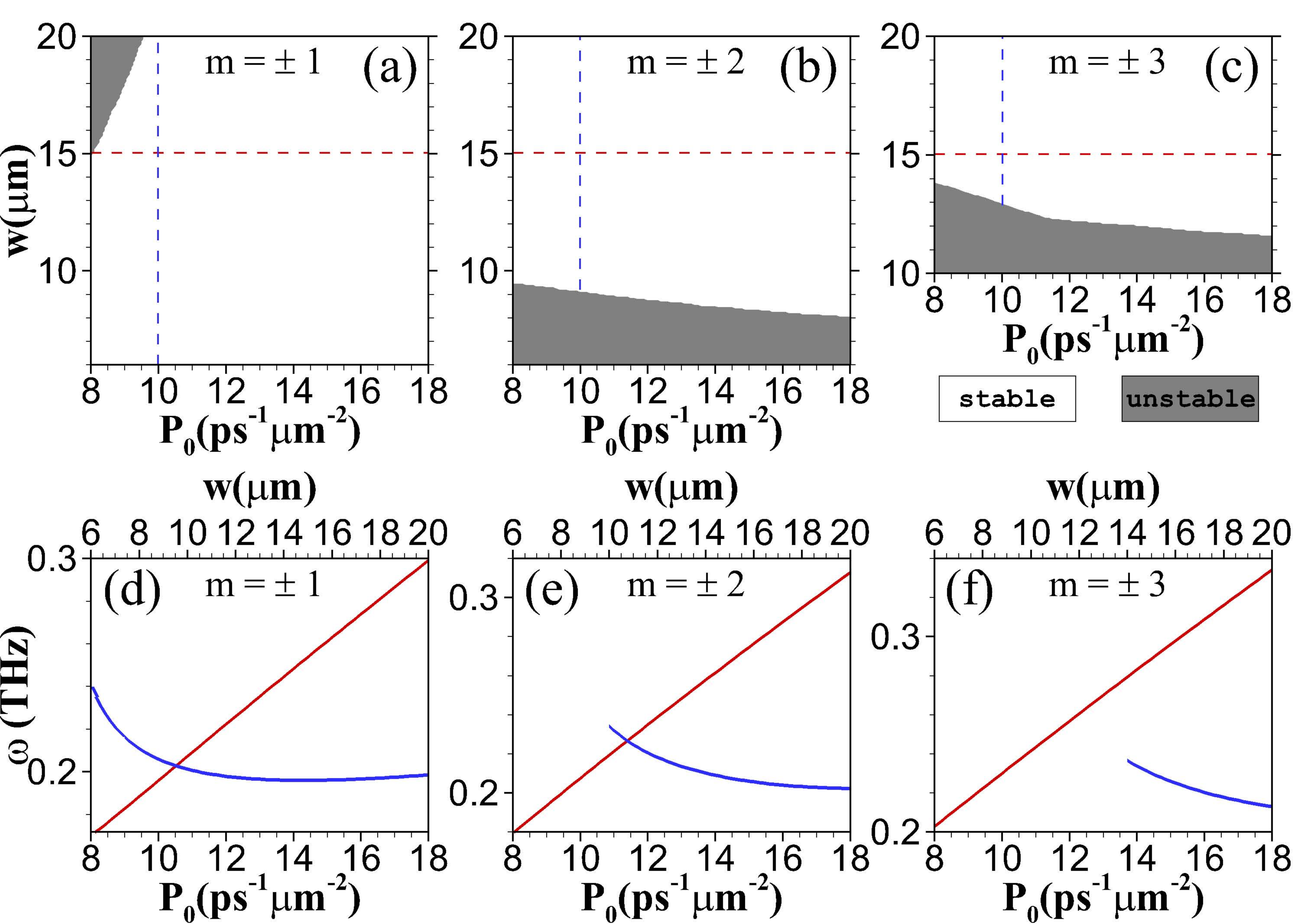}
\caption{(Color online) Stability and instability regions of vortices with (a) $m=\pm1$, (b) $m=\pm2$, and (c) $m=\pm3$, depending on the intensity and the radius of the pump. (d)-(f) Dependence of vortex frequency on the pump intensity (red lines) and pump radius (blue lines). Red lines in (d)-(f) correspond to the red dashed lines in (a)-(c) for a constant pump radius $w=15$ $\mu$m, while blue lines correspond to the blue dashed lines for a constant pump intensity $P_0=10$ $\mathrm{ps^{-1}\mu m^{-2}}$.}\label{stability}
\end{figure}

The numerical stability region of each vortex is shown in Figs.~\ref{stability}(a)-\ref{stability}(c). The stability depends not only on the pump intensity, but also on the pump radius. For vortices with winding number $|m|=1$, unstable solutions appear when the pump radius is too large or the pump intensity is too small [Fig.~\ref{stability}(a)]. When the pump radius is larger, vortices with larger winding number dominate, which is shown in Figs.~\ref{stability}(b) and \ref{stability}(c). While for larger pump intensity the vortex with $|m|=1$ is stabilized because of the stronger confinement potential, vortices with $|m|=2$ and $|m|=3$ become unstable at smaller $w$ when the vortex radius is larger than the pump ring radius. For the same pump radius, the vortices with $|m|=2$ and $|m|=3$ are more robust at larger pump intensity, where the confinement potential is stronger. Vortices with higher topological charges, $|m|>3$, can also be stable at even larger pump radius. In general, the instability region shifts to larger pump radius as the winding number increases.


Figures \ref{stability}(d)-\ref{stability}(f) show that the vortex frequency increases (blue shifts) monotonously with the pump intensity for fixed pump radius as a result of the repulsive nonlinearity [Figs.~\ref{stability}(d)-\ref{stability}(f)]. However, for fixed pump intensity, as the pump radius increases, the frequency decreases (red shifts), even though also in this case the peak density increases as shown below in Fig.~\ref{1Dprofiles}(a). This is because the vortex tail becomes more pronounced as the pump radius increases, strongly depleting the confinement potential, leading to a red shift of the frequency. As the pump radius increases further, the frequency converges to a certain value where the peak density of the vortices becomes radius-independent and the influence of the tail can be almost neglected.

\begin{figure} 
\includegraphics[width=1\columnwidth]{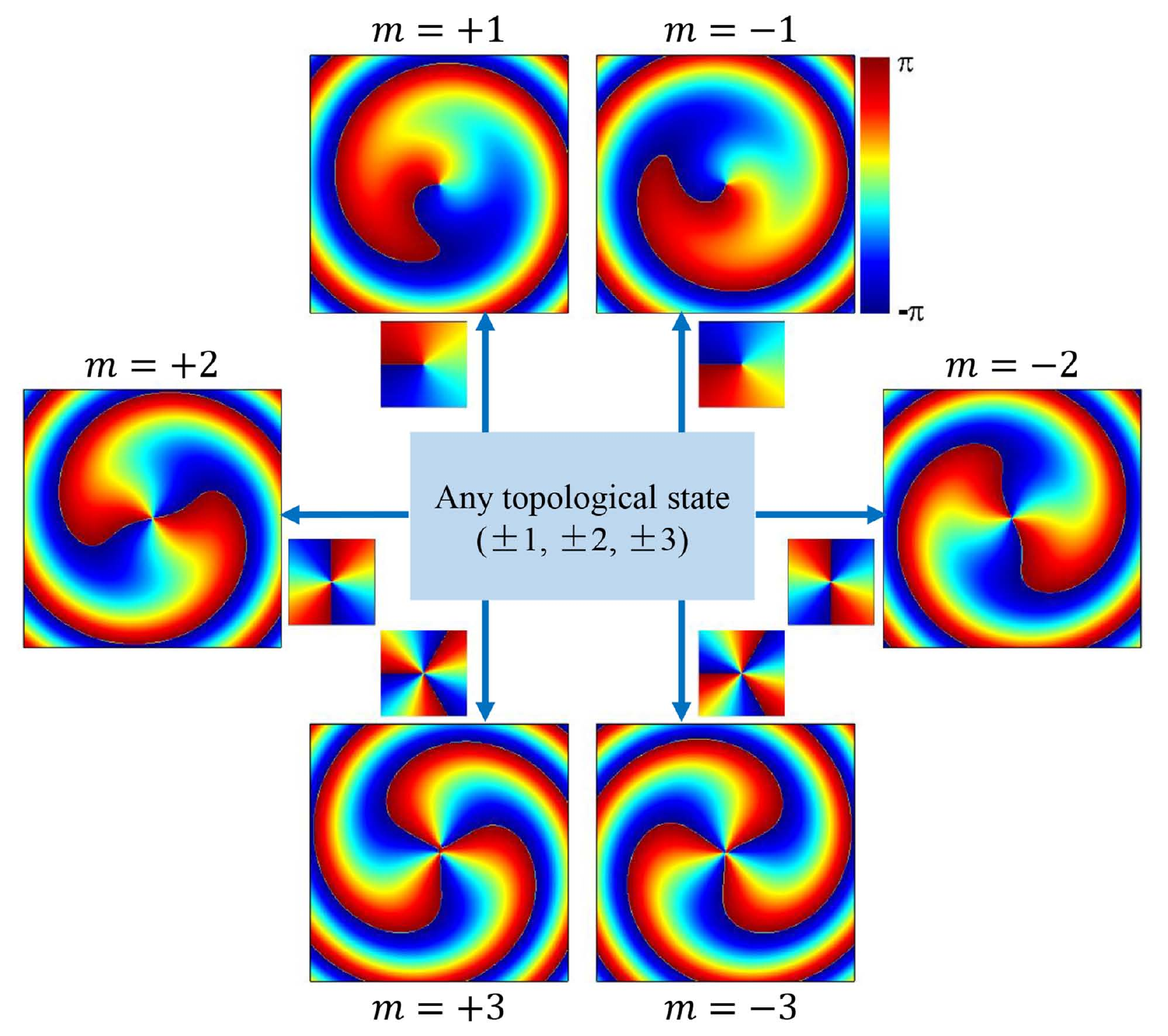}
\caption{(Color online) Illustration of the dynamical switching from any topological state to a desired vortex state. The phase profiles of the target vortex states are shown in the larger panels. The switching between different states is achieved with coherent pulses carrying the orbital angular momentum of the corresponding target state. Phase profiles of the switching pulses are shown in the smaller panels.}\label{switchcoh}
\end{figure}

Now we demonstrate how to address the different vortex solutions individually. Two possible methods come to mind. One includes gradual changes of the pump intensity to approach a bifurcation in a controlled manner. However, in our work there is no obvious pump-intensity dependent bifurcation and hysteresis loop. A pump-radius dependent hysteresis loop does exists around $P_0=8$ $\mathrm{ps^{-1}\mu m^{-2}}$ as is visible in Figs.~\ref{stability}(a) and \ref{stability}(b), however, in an experimental setup it would be difficult to precisely control the pump radius during the excitation. In our system a different approach appears more feasible to switch between different stationary states. This can be achieved by application of additional coherent light pulses as illustrated in Fig.~\ref{switchcoh}. The light pulses carry the same OAM as the target vortex. Ring-shaped pulses are used here,
\begin{equation}\label{cohPulse}
E(\mathbf{r},t)=E_0 {\mathbf{r}^2} e^{-\mathbf{r}^2/w_p^2} e^{-\mathbf{t}^2/w_t^2} e^{im_p\theta} e^{-i\omega_pt},
\end{equation}
with phase profiles shown in Fig.~\ref{switchcoh}. We note that the switching is not very sensitive to the precise choice of pulse amplitude $E_0$, duration $w_t$, radial width $w_p$, and frequency $\omega_p$. To achieve compatibility with the desired target states, here we chose $E_0=0.1$, $w_t=6$ ps, $w_p=10$ $\mu$m, and $\omega_p=0.2$ THz. With these pulses we are able to switch from any vortex state to any desired target state. The pulse profile to be used is determined solely by the final state, the initial state only influences the transient dynamics how the system approaches the new stationary state. For example, an incoherent pulse with $m_p=+1$ leads to a final state with a vortex carrying the topological charge $m=+1$. The initial state can be $m=\pm2$, $m=\pm3$, or $m=-1$. Dynamical details of the topological charge transformation are given in Fig.~S1 of the Supplemental Material (SM). 

\begin{figure} 
\includegraphics[width=1\columnwidth]{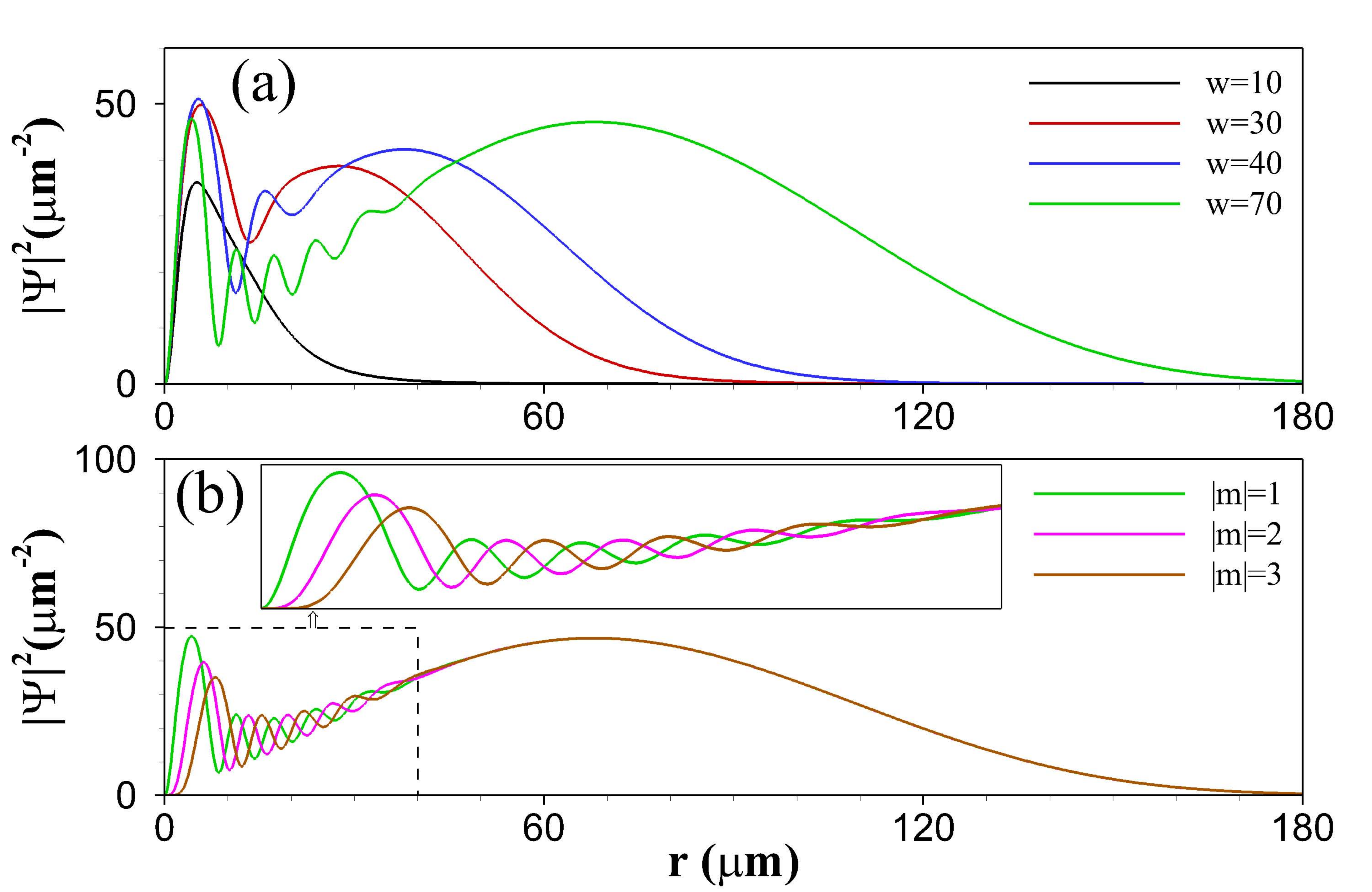}
\caption{(Color online) 1D profiles of vortices with (a) $|m|=1$ for different pump radii at $P_0=15$ $\mathrm{ps^{-1}\mu m^{-2}}$ and (b) different topological charges under the same pump with $w=70$ $\mu$m and $P_0=15$ $\mathrm{ps^{-1}\mu m^{-2}}$.}\label{1Dprofiles}
\end{figure}

\textit{Bessel Vortex} -- The phase of a vortex in a driven-dissipative system has a helical profile~\cite{Crasovan-pre-2000,Lobanov-OL-2011}, which is different from that of a vortex in a conservative system~\cite{Boyd-Book-2003}. In the high-density region of the vortex, however, their phase distributions are similar. Comparing the phases of the different vortex states in Fig. \ref{profiles}, for the $m=+1$ state the phase appears to have a more pronounced radial dependence (it appears to spiral around the center even in the high density region of the vortex). Figure~\ref{1Dprofiles}(a) shows the density cross sections of vortices with $|m|=1$ for different pump radii. When the pump radius is larger, $w=30$ $\mu$m for instance, a broader peak with a longer tail is found. When the pump radius becomes even larger, the broad peak moves far away from the vortex center. Simultaneously, its peak density increases, approaching the peak density of the main peak at small radius. In addition, several additional small peaks appear between the main peak and the broad peak. The corresponding two-dimensional profiles are shown in Figs S2(a)-S2(f) of the SM. While the amplitude of the main peaks is squeezed as the pump radius increases, the radial position of this peak stays almost fixed for the same topological charge [Fig.~\ref{1Dprofiles}(a)]. As shown in Figs.~\ref{1Dprofiles}(b), for different topological charges for the same pump, the radial position of the main peak increases with the topological charge, while the tails of the density almost exactly coincide at larger radii.

\begin{figure} 
\includegraphics[width=1\columnwidth]{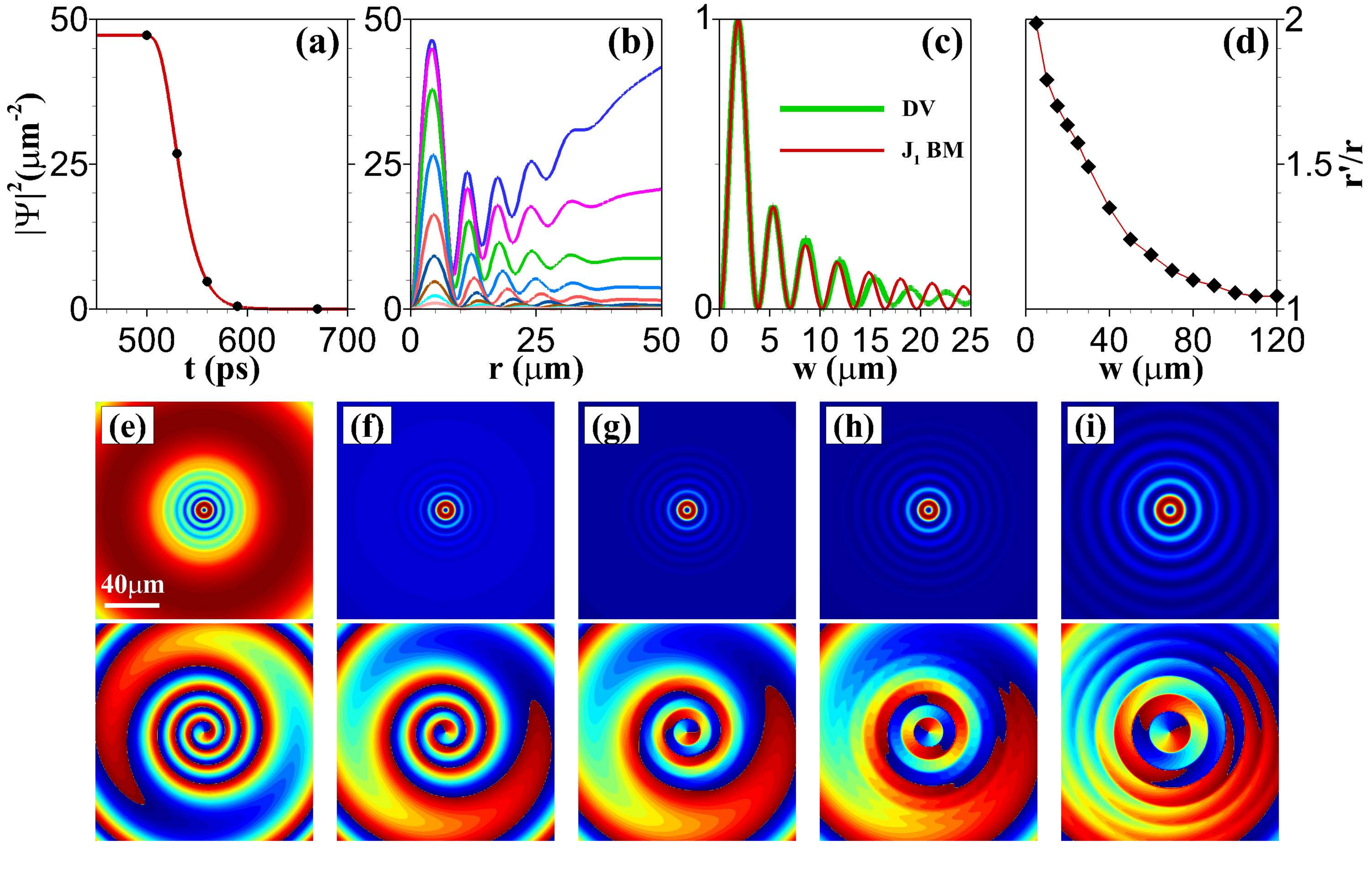}
\caption{(Color online) (a) Time evolution of the peak density of a vortex with $m=1$. The pump is switched off at $t=500$ ps. Pump parameters are $P_0=15$ $\mathrm{ps^{-1}\mu m^{-2}}$ and $w=70$ $\mu$m. (b) 1D profiles of the vortex at different decay time, corresponding, from top to bottom, to $t=500-600$ ps with $10\,\mathrm{ps}$ time interval. (c) Comparison of 1D normalized density profiles of the decayed vortex (DV) at $t=670$ ps and a $J_1$ Bessel mode (BM). (d) Dependence of the ratio $r'/r$ on pump radius. (e)-(i) Profiles of densities and phases of the vortex at different decay time, corresponding to the black points from left to right in (a).}\label{decaym1w70}
\end{figure}

For the vortices with twisted phase profile, we now study the decay dynamics after suddenly switching off the pump. For an only slightly twisted vortex excited by a spatially narrow pump with $w=5$ $\mu$m, after switching off the pump, the vortex radius increases quickly, approaching the double of the original radius as illustrated in Figs.~\ref{decaym1w70}(d) (Details are given in Figs.~S3(a)-S3(e) of the SM). Here, we define $r'$ as the vortex radius when the peak density reduces to one tenth.

For a vortex with strongly twisted phase, however, the decay dynamics is quite different. Figure~\ref{decaym1w70} shows results for $w=70$ $\mu$m. The density decreases to one tenth about 60 ps after switching off the pump [Fig. \ref{decaym1w70}(a)]. Remarkably, during this initial decay the radius of the main peak increases only to 1.13 times the original radius as shown in Figs.~\ref{decaym1w70}(d)-\ref{decaym1w70}(g). The density of the broad peak at larger radius decays much faster than that of the main peak as shown in Fig.~\ref{decaym1w70}(b). As the broad peak decreases over time, the smaller peaks at intermediate radii change to independent features, forming multiple rings outside the main peak when the system approaches the linear regime. The concentric rings are similar to the $J_1$ Bessel mode. The comparison of the profiles of the decayed vortex at $t=670$ ps in the almost linear regime, where the peak density has reduced to $10^{-4}$ of the original peak density and the $J_1$ Bessel mode is shown in Fig.~\ref{decaym1w70}(c). The four main peaks at small radii fit very well, while the difference becomes more evident from the fifth peak onwards. However, in a good approximation, after the initial decay, the vortex formed can be identified as a Bessel vortex. During the decay from the nonlinear to linear regime, the strongly twisted phase evolves from a spiralling smooth curve [Fig. \ref{decaym1w70}(e)] to a distribution with $\pi$ phase jumps [Fig. \ref{decaym1w70}(i)]. We note that the Bessel vortex shows remarkable self-stabilization ability even at strongly reduced densities. Its radius increases much more slowly [Figs. \ref{decaym1w70}(h) and \ref{decaym1w70}(i)] than for the normal vortex at $w=5$ $\mu$m in Fig.~\ref{decaym1w70}(d) [cf. Figs. S3(f)-S3(h)]. This finding is in agreement with recent experiments comparing diffraction of Gaussian vortex beams and Bessel vortex beams in optical fibers~\cite{Ruan-IEEEPJ-2017}. Figure~\ref{decaym1w70}(d) shows that the ratio converges to 1 for increasing pump radius. Obviously, the strongly twisted vortex contains more concentric rings, which is closer to a Bessel mode comparing to the slightly twisted vortex containing only one ring. Further information on the decay dynamics of vortices with higher winding number is given in Fig.~S4 of the SM. 

\textit{Conclusion} -- We report on the existence of a vortex multistability in a polariton condensate where the same incoherent pump can support several stable vortex states. This peculiar behaviour of the driven-dissipative nonlinear system studied is rooted in the intrinsic feedback of the condensate and the excitation reservoir. We demonstrate that coherent light pulses with different OAM can be used to switch between different vortex states. For larger radius of the ring-shaped pump beam, we also find the existence of a Bessel vortex mode, which shows a remarkable persistence even without the pump source.

This work was supported by the Deutsche Forschungsgemeinschaft (DFG) through the collaborative research center TRR142 (project A04) and Heisenberg program (grant No. 270619725) and by the Paderborn Center for Parallel Computing, PC$^2$.


\begin{thebibliography}{99}

\bibitem{Sessoli-Nature-1993}
R. Sessoli, D. Gatteschi, A. Caneschi, and M. A. Novak, Nature {\bf{365}}, 141-143 (1993).

\bibitem{Fujita-Science-1999}
W. Fujita and K. Awaga, Science {\bf{286}}, 261-262 (1999).

\bibitem{Gibbs-APL-1979}
H. M. Gibbs, S. L. McCall, T. N. C. Venkatesan, A. C. Gossard, A. Passner, and W. Wiegmann, Appl. Phys. Lett. {\bf{35}}, 451 (1979).

\bibitem{Baas-PRA-2004}
A. Baas, J. Ph. Karr, H. Eleuch, and E. Giacobino, Phys. Rev. A {\bf{69}}, 023809 (2004).

\bibitem{Treutlein-PRL-2007}
P. Treutlein, D. Hunger, S. Camerer, T. W. H\"{a}nsch, and Jakob Reichel, Phys. Rev. Lett. {\bf{99}}, 140403 (2007).

\bibitem{Soljacic-PRE-2002}
M. Solja{\v{c}}i{\'c}, M. Ibanescu, S. G. Johnson, Y. Fink, and J. D. Joannopoulos, Phys. Rev. E {\bf{66}}, 055601 (2002).

\bibitem{Boyd-Book-2003}
R. W. Boyd, \textit{Nonlinear Optics}, Academic press (2003).

\bibitem{Fraile-OL-1991}
F. J. Fraile-Pel\'{a}ez, J. Capmany, and M. A. Muriel, Opt. Lett. {\bf{16}}, 907-909 (1991).

\bibitem{Lyons-APL-1995}
E. R. Lyons and G. J. Sonek, Appl. Phys. Lett. {\bf{66}}, 1584 (1995).

\bibitem{Yanik-APL-2003}
M. F. Yanik and S. Fan, Appl. Phys. Lett. {\bf{83}}, 2739 (2003).

\bibitem{Notomi-OE-2005}
M. Notomi, A. Shinya, S. Mitsugi, G. Kira, E. Kuramochi, and T. Tanabe, Opt. Exp. {\bf{13}}, 2678-2687 (2005).

\bibitem{Deng-science-2002}
H. Deng, G. Weihs, C. Santori, J. Bloch, and Y. Yamamoto, Science {\bf{298}}, 199 (2002).

\bibitem{Kasprzak-nature-2006}
J. Kasprzak, M. Richard, S. Kundermann, A. Baas, P. Jeambrun, J. M. J. Keeling, F. M. Marchetti, M. H. Szyma\'{n}ska, R. Andr\'{e}, J. L. Staehli, V. Savona, P. B. Littlewood, B. Deveaud, and Le Si Dang, Nature {\bf{443}}, 409 (2006).

\bibitem{Christopoulos-2007-prl}
S. Christopoulos, G. Baldassarri H\"{o}ger von H\"{o}gersthal, A. J. D. Grundy, P. G. Lagoudakis, A.V. Kavokin, J. J. Baumberg, G. Christmann, R. Butt\'{e}, E. Feltin, J.-F. Carlin, and N. Grandjean, Phys. Rev. Lett. \textbf{\bf 98}, 126405 (2007).

\bibitem{Christmann-2008-apl}
G. Christmann, R. Butt\'{e}, E. Feltin, J.-F. Carlin, and N. Grandjean, Appl. Phys. Lett. \textbf{\bf 93}, 051102 (2008).

\bibitem{Baumberg-2008-prl}
J. J. Baumberg, A. V. Kavokin, S. Christopoulos, A. J. D. Grundy, R. Butt\'{e}, G. Christmann, D. D. Solnyshkov, G. Malpuech, G. Baldassarri H\"oger von H\"ogersthal, E. Feltin, J.-F. Carlin, and N. Grandjean, Phys. Rev. Lett. \textbf{\bf 101}, 136409 (2008).

\bibitem{Plumhof-2013-nm}
J. D. Plumhof, T. St\"{o}ferle, L. Mai, U. Scherf, and R. F. Mahrt, Nat. Mater. {\bf{13}}, 247-252 (2013).

\bibitem{Borgh-PRB-2010}
M. O. Borgh, J. Keeling, and N. G. Berloff, Phys. Rev. B {\bf{81}}, 235302 (2010).

\bibitem{Luk-PRB-2013}
M. H. Luk, Y. C. Tse, N. H. Kwong, P. T. Leung, P. Lewandowski, R. Binder, and S. Schumacher, Phys. Rev. B {\bf{87}}, 205307 (2013).

\bibitem{Werner-PRB-2014}
A. Werner, O. A. Egorov, and F. Lederer, Phys. Rev. B {\bf{89}}, 245307 (2014).

\bibitem{Liew-PRB-2015}
T. C. H. Liew, O. A. Egorov, M. Matuszewski, O. Kyriienko, X. Ma, and E. A. Ostrovskaya, Phys. Rev. B {\bf{91}}, 085413 (2015).

\bibitem{Egorov-2009-prl}
O. A. Egorov, D. V. Skryabin, A. V. Yulin, and F. Lederer, Phys. Rev. Lett. \textbf{\bf 102}, 153904 (2009).

\bibitem{Sich-2012-nph}
M. Sich, D. N. Krizhanovskii, M. S. Skolnick, A. V. Gorbach, R. Hartley, D. V. Skryabin, E. A. Certa-M\'{e}ndez, K. Biermann, R. Hey, and P. V. Santos, Nat. Photonics \textbf{\bf 6}, 50 (2012).

\bibitem{Yulin-2008-pra}
A. V. Yulin, O. A. Egorov, F. Lederer, and D. V. Skryabin, Phys. Rev. A \textbf{\bf 78}, 061801(R) (2008).

\bibitem{Grosso-2012-prb}
G. Grosso, G. Nardin, F. Morier-Genoud, Y. L\'{e}ger, and B. Deveaud-Pl\'{e}dran, Phys. Rev. B \textbf{\bf 86}, 020509(R) (2012).

\bibitem{Cilibrizzi-2014-prl}
P. Cilibrizzi, H. Ohadi, T. Ostatnicky, A. Askitopoulos, W. Langbein, and P. Lagoudakis, Phys. Rev. Lett. \textbf{\bf 113}, 103901 (2014).

\bibitem{Ma-2017-prl}
X. Ma, O. A. Egorov, and S. Schumacher, Phys. Rev. Lett. \textbf{\bf 118}, 157401 (2017).

\bibitem{Bajoni-PRL-2008}
D. Bajoni, E. Semenova, A. Lema\^{\i}tre, S. Bouchoule, E. Wertz, P. Senellart, S. Barbay, R. Kuszelewicz, and J. Bloch, Phys. Rev. Lett. \textbf{\bf 101}, 266402 (2008).

\bibitem{Liew-PRB-2010}
T. C. H. Liew, A. V. Kavokin, T. Ostatnicky, M. Kaliteevski, I. A. Shelykh, and R. A. Abram, Phys. Rev. B \textbf{\bf 82}, 033302 (2010).

\bibitem{Ballarini-NC-2013}
D. Ballarini, M. De Giorgi, E. Cancellieri, R. Houdr\'{e}, E. Giacobino, R. Cingolani, A. Bramati, G. Gigli, and D. Sanvitto, Nat. Commun. \textbf{\bf 4}, 1778 (2013).

\bibitem{Kyriienko-PRL-2014}
O. Kyriienko, T. C. H. Liew, and I. A. Shelykh, Phys. Rev. Lett. \textbf{\bf 112}, 076402 (2014).

\bibitem{Tan-PRB-2018}
E. Z. Tan, H. Sigurdsson, and T. C. H. Liew, Phys. Rev. B \textbf{\bf 97}, 075305 (2018).

\bibitem{Lagoudakis-NatPhys-2008}
K. G. Lagoudakis, M. Wouters, M. Richard, A. Baas, I. Carusotto, R. Andr\'{e}, Le Si Dang, and B. Deveaud-Pl\'{e}dran, Nat. Phys. {\bf{4}}, 706 (2008).

\bibitem{Lagoudakis-Science-2009}
K. G. Lagoudakis, T. Ostatnicky, A. V. Kavokin, Y. G. Rubo, R. Andr\'{e}, B. Deveaud-Pl\'{e}dran, Science {\bf{326}}, 974 (2009).

\bibitem{Sigurdsson-prb-2014}
H. Sigurdsson, O. A. Egorov, X. Ma, I. A. Shelykh, and T. C. H. Liew, Phys. Rev. B {\bf{90}}, 014504 (2014).

\bibitem{Ma-prb-2016}
X. Ma, U. Peschel, and O. A. Egorov, Phys. Rev. B {\bf{93}}, 035315 (2016).

\bibitem{Dall-PRL-2014}
R. Dall, M. D. Fraser, A. S. Desyatnikov, G. Li, S. Brodbeck, M. Kamp, C. Schneider, S. H\"{o}fling, and E. A. Ostrovskaya, Phys. Rev. Lett. {\bf{113}}, 200404 (2014).

\bibitem{Ma-prb-2017}
X. Ma and S. Schumacher, Phys. Rev. B {\bf{95}}, 235301 (2017).

\bibitem{Schmutzler-prb-2015}
J. Schmutzler, P. Lewandowski, M. A{\ss}mann, D. Niemietz, S. Schumacher, M. Kamp, C. Schneider, S. H\"{o}fling, and M. Bayer, Phys. Rev. B {\bf{91}}, 195308 (2015).

\bibitem{Chong-NP-2010}
A. Chong, W. H. Renninger, D. N. Christodoulides, and F. W. Wise, Nat. Photonics {\bf{4}}, 103-106 (2010).

\bibitem{Alexandrescu-PRA-2006}
A. Alexandrescu and V. M. P\'{e}rez-Garc\'{\i}a, Phys. Rev. A {\bf{73}}, 053610 (2006).

\bibitem{Wouters-prl-2007}
M. Wouters and I. Carusotto, Phys. Rev. Lett. {\bf{99}}, 140402 (2007).

\bibitem{Roumpos-NatPhys-2011}
G. Roumpos, M. D. Fraser, A. L\"{o}ffler, S. H\"{o}fling, A. Forchel, and Y. Yamamoto, Nat. Phys. {\bf{7}}, 129 (2011).

\bibitem{Deng-PNAS-2003}
H. Deng, G. Weihs, D. Snoke, J. Bloch, and Y. Yamamoto, Proc. Natl. Acad. Sci. \textbf{\bf 100}, 15318-15323 (2003).

\bibitem{Love-PRL-2008}
A. P. D. Love, D. N. Krizhanovskii, D. M. Whittaker, R. Bouchekioua, D. Sanvitto, S. A. Rizeiqi, R. Bradley, M. S. Skolnick, P. R. Eastham, R. Andr\'{e}, and L. S. Dang, Phys. Rev. Lett. \textbf{\bf 101}, 067404 (2008).

\bibitem{Wertz-NP-2010}
E. Wertz, L. Ferrier, D. D. Solnyshkov, R. Johne, D. Sanvitto, A. Lema\^{\i}tre, I. Sagnes, R. Grousson, A. V. Kavokin, P. Senellart, G. Malpuech, and J. Bloch, Nat. Phys. \textbf{\bf 6}, 860-864 (2010).

\bibitem{Nelsen-PRX-2013}
B. Nelsen, G. Liu, M. Steger, D. W. Snoke, R. Balili, K. West, and L. Pfeiffer, Phys. Rev. X \textbf{\bf 3}, 041015 (2013).

\bibitem{Steger-Optica-2015}
M. Steger, C. Gautham, D. W. Snoke, L. Pfeiffer, and K. West, Optica \textbf{\bf 2}, 1-5 (2015).

\bibitem{Sun-PRL-2017}
Y. Sun, P. Wen, Y. Yoon, G. Liu, M. Steger, L. N. Pfeiffer, K. West, D. W. Snoke, and K. A. Nelson, Phys. Rev. Lett. \textbf{\bf 118}, 016602 (2017).

\bibitem{Ferrier-PRL-2011}
L. Ferrier, E. Wertz, R. Johne, D. D. Solnyshkov, P. Senellart, I. Sagnes, A. Lema\^{\i}tre, G. Malpuech, and J. Bloch, Phys. Rev. Lett. \textbf{\bf 106}, 126401 (2011).

\bibitem{Sanvitto-NM-2016}
D. Sanvitto and S. K\'{e}na-Cohen, Nat. Mater. {\bf{15}}, 1061-1073 (2016).

\bibitem{Daskalakis-PRL-2015}
K. S. Daskalakis, S. A. Maier, and S. K\'{e}na-Cohen, Phys. Rev. Lett. \textbf{\bf 115}, 035301 (2015).

\bibitem{Lerario-NP-2017}
G. Lerario, A. Fieramosca, F. Barachati, D. Ballarini, K. S. Daskalakis, L. Dominici, M. De Giorgi, S. A. Maier, G. Gigli, S. K\'{e}na-Cohen, and D. Sanvitto, Nat. Phys. \textbf{\bf 13}, pages 837-841 (2017).

\bibitem{Crasovan-pre-2000}
L.-C. Crasovan, B. A. Malomed, and D. Mihalache, Phys. Rev. E {\bf{63}}, 016605 (2000).

\bibitem{Lobanov-OL-2011}
V. E. Lobanov, Y. V. Kartashov, V. A. Vysloukh, and L. Torner, Opt. Lett. {\bf{36}}, 85-87 (2011).

\bibitem{Ruan-IEEEPJ-2017}
H. Ruan, L. Wang, S. Wu, B. Zhou, IEEE Photon. J. {\bf{9}}, 6500910 (2017).


\end{thebibliography}
\end{document}